\documentclass[%
 aip,
 amsmath,amssymb,
 reprint,%
]{revtex4-1}

\usepackage{graphicx}% Include figure files
\usepackage{dcolumn}% Align table columns on decimal point
\usepackage{bm}% bold math
\usepackage[utf8]{inputenc}
\usepackage[T1]{fontenc}
\usepackage{mathptmx}

\usepackage[ngerman]{babel}
\usepackage{siunitx}
\usepackage{caption}
\usepackage{subcaption}
\usepackage{color}
\usepackage[usenames,dvipsnames]{xcolor}
\usepackage{hyperref}
\usepackage{booktabs}
\usepackage{upgreek}
\usepackage{setspace}
\graphicspath{{img/}}															% legt Unterverzeichnisse fest in denen \includegraphics standardmäßig nach Bilddateien sucht
\hypersetup{ %
	colorlinks = false,		% true: stellt Verknüpfungen wie Links, Zitate und Verweise farbig dar, false: rahmt Links, Zitate und Verweise in einen farbigen Rahmen ein
	urlcolor = darkgray,		% Schriftfarbe für URLs
	citecolor = darkgray,		% Schriftfarbe für Zitate
	pdfborder = 0 0 1,		% legt Rahmen fest, sofern verwendet
	linkbordercolor = white,	% Rahmenfarbe bei Verweisen (white = transparent bei weißem Hintergrund)
	urlbordercolor = black,		% Rahmenfarbe bei URLs
	citebordercolor = white,		% Rahmenfarbe bei Zitaten (white = transparent bei weißem Hintergrund)
	bookmarksopen = false,		% legt fest ob die Lesezeichenleiste beim Öffnen des PDFs expandiert ist oder nicht
	bookmarksnumbered = false,	% legt fest ob die Kapitelnummern in den Lesezeichenbaum übernommen werden
	}

\newcommand{\unit}[1]{\ensuremath{\, \mathrm{#1}}}

	\let\olditemize\itemize
	\renewcommand{\itemize}{
		\olditemize
		\itemsep4pt
	}

	\let\oldenumerate\enumerate
	\renewcommand{\enumerate}{
		\oldenumerate
		\itemsep4pt
	}

\addto\extrasngerman{}
\addto\extrasngerman{}
\addto{\captionsngerman}{%

}

\newcommand{\gka}[1]{}
\usepackage{perpage}

\newcommand{\red}{} 
\newcommand{\blue}{} 
\newcommand{\green}{} 

\newcommand{\bs}[1]{}
\newcommand{\hs}[1]{}
\newcommand{\dl}[1]{}

\begin{document}
\pagenumbering{arabic}

\title{Versatile high-speed confocal microscopy using a single laser beam\\}

\author{Benedikt B. Straub}
\affiliation{ 
Max Planck Institute for Polymer Research, Ackermannweg 10, D-55128 Mainz, Germany%\\This line break forced with \textbackslash\textbackslash
}%
\author{David C. Lah}%
\affiliation{ 
Max Planck Institute for Polymer Research, Ackermannweg 10, D-55128 Mainz, Germany%\\This line break forced with \textbackslash\textbackslash
}%
\author{Henrik Schmidt}%
\affiliation{ 
Max Planck Institute for Polymer Research, Ackermannweg 10, D-55128 Mainz, Germany%\\This line break forced with \textbackslash\textbackslash
}%
\author{Marcel Roth}%
\affiliation{ 
Max Planck Institute for Polymer Research, Ackermannweg 10, D-55128 Mainz, Germany%\\This line break forced with \textbackslash\textbackslash
}%
\author{Laurent Gilson}%
\affiliation{ 
Max Planck Institute for Polymer Research, Ackermannweg 10, D-55128 Mainz, Germany%\\This line break forced with \textbackslash\textbackslash
}%
\author{Hans-Jürgen Butt}%
\affiliation{ 
Max Planck Institute for Polymer Research, Ackermannweg 10, D-55128 Mainz, Germany%\\This line break forced with \textbackslash\textbackslash
}%
\author{Günter K. Auernhammer}%
\email{auernhammer@ipfdd.de}
\affiliation{ 
Max Planck Institute for Polymer Research, Ackermannweg 10, D-55128 Mainz, Germany%\\This line break forced with \textbackslash\textbackslash
}%
\affiliation{ 
Leibnitz Institute for Polymer Research, Hohe Stra\ss e 6, D-01069 Dresden, Germany%\\This line break forced with \textbackslash\textbackslash
}%

\begin{abstract}
We present a new flexible high speed laser scanning confocal microscope and its extension by an astigmatism particle tracking device (APTV). 
Many standard confocal microscopes use either a single laser beam to scan the sample at relatively low overall frame rate, or many laser beam to simultaneously scan the sample and achieve a high overall frame rate. 
Single-laser-beam confocal microscope often use a point detector to acquire the image. 
To achieve high overall frame rates, we use, next to the standard 2D probe scanning unit, a second 2D scan unit projecting the image directly on a 2D CCD-sensor (re-scan configuration).
Using only a single laser beam eliminates cross-talk and leads to an imaging quality that is independent of the frame rate with a lateral resolution of 0.235\unit{\mu m}. 
The design described here is suitable for high frame rate, i.e., for frame rates well above video rate (full frame) up to a line rate of 32kHz.
The dwell time of the laser focus on any spot in the sample (122ns) is significantly shorter than in standard confocal microscopes (in the order of milli or microseconds). 
This short dwell time reduces phototoxicity and bleaching of fluorescent molecules.
The new design opens further flexibility and facilitates coupling to other optical methods. 
The setup can easily be extended by an APTV device to measure three dimensional dynamics while being able to show high resolution confocal structures. 
Thus one can use the high resolution confocal information synchronized with an APTV dataset.
\end{abstract}

\date{\today}

\maketitle

\section{Introduction}

In conventional fluorescence microscopy, up to 90\% of the observed fluorescence can originate from parts of the sample that are out of focus.
This unspecific background from above and below the area-of-interest may lead to high unspecific background noise in the images and overall faint images of the sample. \cite{conchello2005}
Confocal microscopy, introduced by Minsky in 1957, \cite{minsky_88}  is known to solve this problem by blocking all out of focus information with a pinhole. 
Since its invention laser scanning confocal microscopy (in which the confocal detection volume is scanned through the sample) has become a standard tool in many fields of science.
Examples include biological and bio-medical applications, \cite{pawley2006book, grewe2010,  prevedel16, shtrahman15, carrillo-reid16, yang18} 
%{\blue \cite{ har-gil18, takahashi2011}}
 material science, \cite{hovis2010} soft matter science \cite{weitz06a,leunissen07b,jenkins2011, kurita2012,taffs2013,ghosh2010prl,hsiao17, kamp16, meijer18, roller18}
% {\blue \cite{ sujit2013, weitz06a, elmasri2012, leunissen07b, jenkins2011, pham2006, kurita2012,hunter2011,taffs2013,ghosh2010prl}}
and wetting applications. \cite{schellenberger18,bazazi19}
In many of these applications, a major limiting factor is the acquisition time needed in 3D imaging.

The need for high-speed 3D imaging has led to various strategies to increase the scanning speed.
Among them are several varieties of multi-beam scanning confocal microscopes, spinning disc, \cite{nakano2002,shimozawa2013,toomre2006} pinhole array setups \cite{kagawa2013} and slit scanning microscopes. \cite{castellano2012,vogt2010,sabharwal1999}
Especially in thick samples with a high concentration of fluorescent objects, these multi-beam setups lead to a considerable background noise, originating from cross talk between the beams. \cite{graef2005, winter14}
Moreover out of focus photons still contribute to phototoxicity and photodamage.
Therefore samples should be resistive to photodamage when using these methods. 
This can be overcome by using two photon excitation \cite{shimozawa2013,packer15,rickgauer14} or, at lower resolutions, laser light sheet microscopy. \cite{mertz11,chen14,power17,royer18}

When imaging at high frame rates, moving the focal plane without disturbing the system is challenging.
Various methods to move the focal plane to different illumination depths to enable high speed 3D imaging were developed.
A spatial light modulator (SLM) can be used to generate holographic light patterns to scan the whole volume of the probe. \cite{shtrahman15, yang18, packer15, nikolenko08}
{\blue Another method is a tunable lens like an ultrasound lens enabling volumetric scanning with tens of hertz \cite{kong15} or liquid lenses.\cite{duocastella_14, piazza_18} }
A different technique is remote focusing. \cite{botcherby08, anselmi11, botcherby12, curran14}
For this application, additional objectives are placed in the beam path.
In one case, a fast movable mirror is placed in the the focal plane of an additional objective. 
The incoming light is reflected by the mirror into the pupil plane of the imaging objective.
Due to the movement of the mirror, the focal spot can be scanned in z-direction (vertical to the sample plane) through the sample. \cite{botcherby12}
In a different configuration two additional objectives are placed head to head in the beampath.
If one of the objectives is moved, the focal spot is also moved through the sample. \cite{curran14}

\begin{figure*}[tbp]
	\begin{center}
	\includegraphics[width=0.65\textwidth]{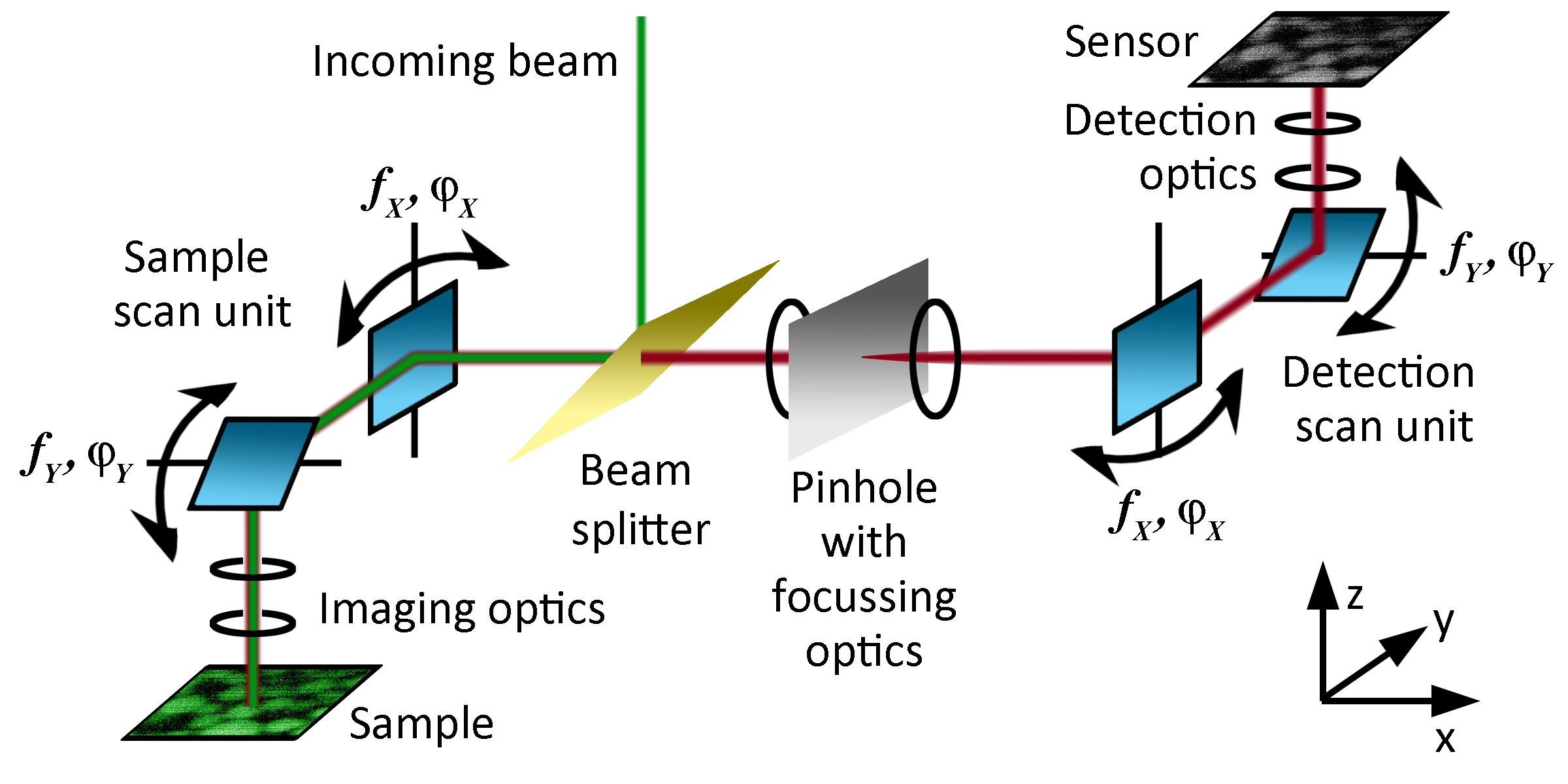}
	\caption{Schematic beam path of the microscope. Shown is a simplified version of the entire beam path, showing symmetry between the sample scanning and the detection scanning as the key element of the present setup. Despite this symmetry, all scanning mirrors can be addressed individually, allowing for a free choice of the scanning amplitudes (within a certain limit for the amplitude ratio) while keeping the frequency and phase coupling effective. Both scanning units are identical. Fast resonant mirrors ($f_r = 16 \unit{kHz}$) give a fixed line frequency of $f_x = 32 \unit{kHz}$ and an adjustable amplitude. The $y$-axes is realised by galvanic mirrors with freely adjustable frequency $f_y$ and amplitude. 
	}
	\label{fig:scheme-path}
	\end{center}
\end{figure*}

Fast 3D scanning, like in the examples above, aims at imaging the entire 3D volume at high frame rate. 
In some cases, however, 3D information on only a portion of the objects in the sample is sufficient. 
Astigmatism particle tracking velocimetry (APTV) \cite{kao94, cierpka11, rossi14} is another method without the need of moving optical objects to measure three dimensional dynamics.
In this method, a cylindrical element is placed in the optical path of a fluorescent microscope directly in front of the CCD chip.
This cylindrical element creates an aberration (astigmatism) of the image. 
The images of spherical fluorescent tracer particles are deformed to ellipsoids of which the aspect ratio is a function of the position of the tracer particle along the beam direction.
The shape information is used to calculate the 3D position of the particle.
Therefore this method can calculate 3D information out of a 2D measurement without the need of moving optical elements.
This simplifies the optical assembling, increases the frame rate and widens the observable volume at the expense of losing resolution.

In this article we describe a laser scanning confocal microscope setup that combines the advantages of high scanning speeds and the use of a single scanning beam.
We use two independent but synchronized scanning units.
Using a single beam eliminates cross-talk and lead to an imaging quality that is independent of the frame rate.
An astigmatic configuration of the setup is explained in the later sections which enables easy measurements of three dimensional dynamics.
The present setup is an extension to the homebuilt microscope that was described in previous publications. \cite{roth2012b,roth2012a}
Some ideas of the present setup have been published in Auernhammer et al..\cite{patent}
Whereas Auernhammer\cite{patent} focused on the basic principles of such design, here, we aim to demonstrate the feasibility and implementation and characterize the possibilities. 
Our new design with two independent but synchronized scanning units opens further flexibility, as we will illustrate below.
\gka{Dem kritischen Referee müssen wir in diesem Abschnitt schon ein Gegenargument liefern. Warum ist so eine Entwicklung notwendig?  Ein mögliches Argument wäre, dass man Strukturaufklärung und 3D Trajektorien mit > 100 Hz braucht um Fließen von granularen Suspensionen zu verstehen (SFB 1194 FP 2 und das Projekt von Himanshu). Normale schnelle konfokale Mikroskopie schafft Trajektorien nur mit der 3D-Bildrate. Wir wollen aber schneller sein. Also splitten wir das Problem in zwei Teile: 2D für alle Partikel (schnelle konfokale Mikroskopie) und 3D für eine repräsentative Untermenge (aPTV). Letztere liefert Trajektorien und zur Struktur eine Statistik, erstere keine Trajektorien aber Korrelationen in der Struktur. Der wissenschaftliche Inhalt kommt in einem späteren Paper, hier kommt nur die Methodik, die das ermöglichen soll. }

\section{Experimental setup}
\label{sect:realize}

\begin{figure}[tbp]
	\begin{center}
	\includegraphics[width=\columnwidth]{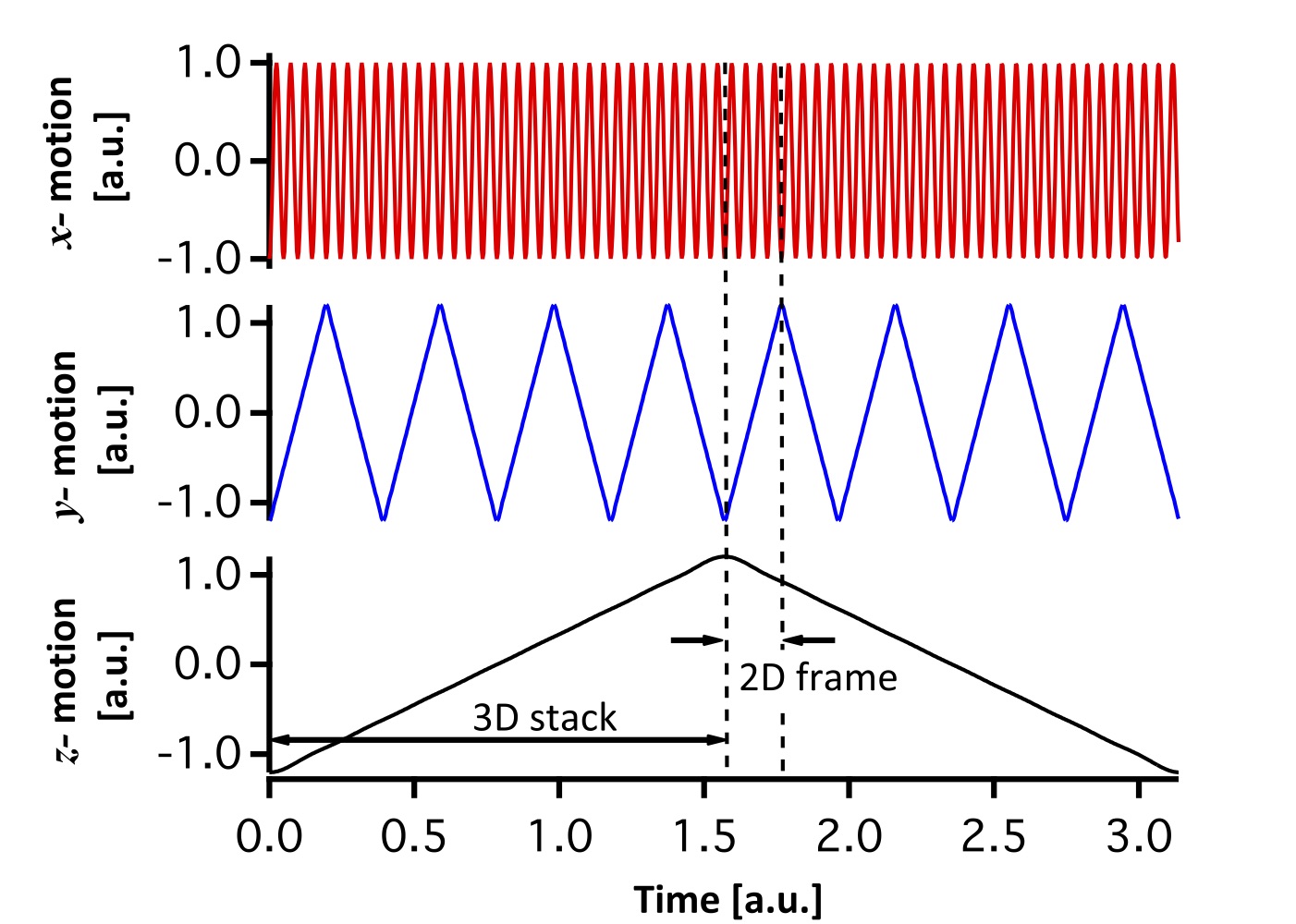}
	\caption{Sketch of the synchronized motion of the scanning units in the case of the standard 3D mode. The resonant mirrors perform a sinusoidal motion, scanning two lines per period of motion. To minimize dead time,  the $y$- and $z$-motion is rectilinear and scanned also in a bidirectional way. The number of lines per 2D frame and the number of 2D frames per 3D stack is adjustable through the frequencies chosen to drive the galvanic mirrors and the $z$ motion of the focus.}
	\label{fig:scanning}
	\end{center}
\end{figure}

The optical path from the light source (incoming beam) to the sample and back through the beam splitter and the pinhole follows the standard design of single beam confocal microscopes (Fig.~\ref{fig:scheme-path}).
As light source we use green laser light (Cobolt Samba 532nm, 100mW, Cobolt AB, Solna, Sweden). 
The incoming beam is led to the sample scan unit by a beam splitter (N-BK7, Qioptiq, Göttingen, Germany) .
The scan unit contains a resonant scanner (SC 30, fixed frequency: 16kHz, EOPC, Ridgewood, NY, USA) and a galvanic scanner (Dynaxis XS, Scanlab, Puchheim, Germany).
Due to the oscillation of the scanners, the scan unit creates a diverging array of laser beams.
A telecentric system in the imaging optics (CLS-SL combined with ITL200, Thorlabs, Munich, Germany) transforms it to a parallel array of laser beams.
Hence, to change the focal plane, we are able to translate the objective (UPLSAPO 60XO, Olympus, Tokio, Japan) along the optical path without loss in optical quality of the beam path.
The $z$-position of the objective is actuated with a piezo-positioning system (PD72Z4CA0, Physik Instrumente, Karlsruhe, Germany) with  a range of 400$\mu m$.
The objective focuses the laser beam onto the sample and collects the fluorescent or reflected light. 
The collected light passes through the beam splitter and is projected onto a 20$\mu m$ pinhole (Qioptiq, Göttingen, Germany). 
Tests with pinholes of different diameters reveal that there is an optimum pinhole diameter between 20$\mu m$ and $50\unit{\mu m}$. 
This diameter balances resolution and collecting efficiency for the light coming from the sample.

In standard single beam confocal microscopes the intensity would be measured after this pinhole by a fast photon counter, e.g., an avalanche diode or a photo-multiplier tube.
Instead, in our setup, we included a second but identical scan unit between the pinhole and the sensor.
Different to other setups that used one scan unit twice, \cite{castellano2012,vogt2010,sabharwal1999} we use two separate 2D scan units that work in a phase-locked (but not amplitude-locked) mode.
Our setup has some similarities to the re-scan confocal microscopes in De Luca et al.. \cite{de13,de17, de17-frAQB}
However, we use not only synchronized galvanometer scanners but also synchronized resonant scanners. 
This allows for higher scanning rates. 
The synchronization between the two scan units is achieved with a single driver (PLD-2S-220 Driver, EOPC, Fresh Meadows, NY, USA) controlling both resonant mirrors and with a common frequency generator feeding the drives of the galvanic mirrors.
This configuration allows for an independent choice of the scanning amplitudes in both scanning units.
For the resonant mirrors the synchronization is only efficient for moderate differences in the amplitudes.
The rescanned beam is focused on the 2D sensor of a high-speed camera (\mbox{1024px x 1024px,} 14$\mu m$ pixel size, MotionXtra NR4-S1, IDT Vision, USA) using an f-theta lens (S4LFT1254/121, Sill Optics, Wendelstein, Germany).
To improve the signal to noise ratio, we enclosed the entire setup to shield it from ambient light.

In standard operation mode, the acquisition of the camera is synchronized with both galvanic mirrors to record a $xy$-plane.
By synchronising the acquisition to both the galvanic mirrors and the piezo-positioning system of the objective, scans of 3D stacks can be accomplished.
However, the two scanning units can also be driven separately from each other, allowing for the projection of other planes in the sample on the 2D sensor.
When the motion of the re-scanning galvanic mirror is synchronized to the piezo-positioning system of the objective, an $xz$-plane can be projected on the sensor.
With a more complex combination of the piezo-positioning system and the sample scanning galvanic mirror, any plane that contains the motion of the resonant mirror can be projected on the 2D sensor.

The resonant mirrors perform a sinusoidal motion, leading to a reduced scanning speed close to the turning points of the beam in $x$-direction.
The motion along the $y$- and $z$-axes is chosen to be of constant speed, i.e., with a symmetric triangular signal of the driving voltage vs.~time (Fig. \ref{fig:scanning}).

\section{Characterization of the optical properties}
\label{sect:optics}
\subsection{Results on reflective calibration patterns}
\label{sect:line_grid}

\begin{figure*}[tbp]
	\begin{center}
	\includegraphics[width = 143.293mm]{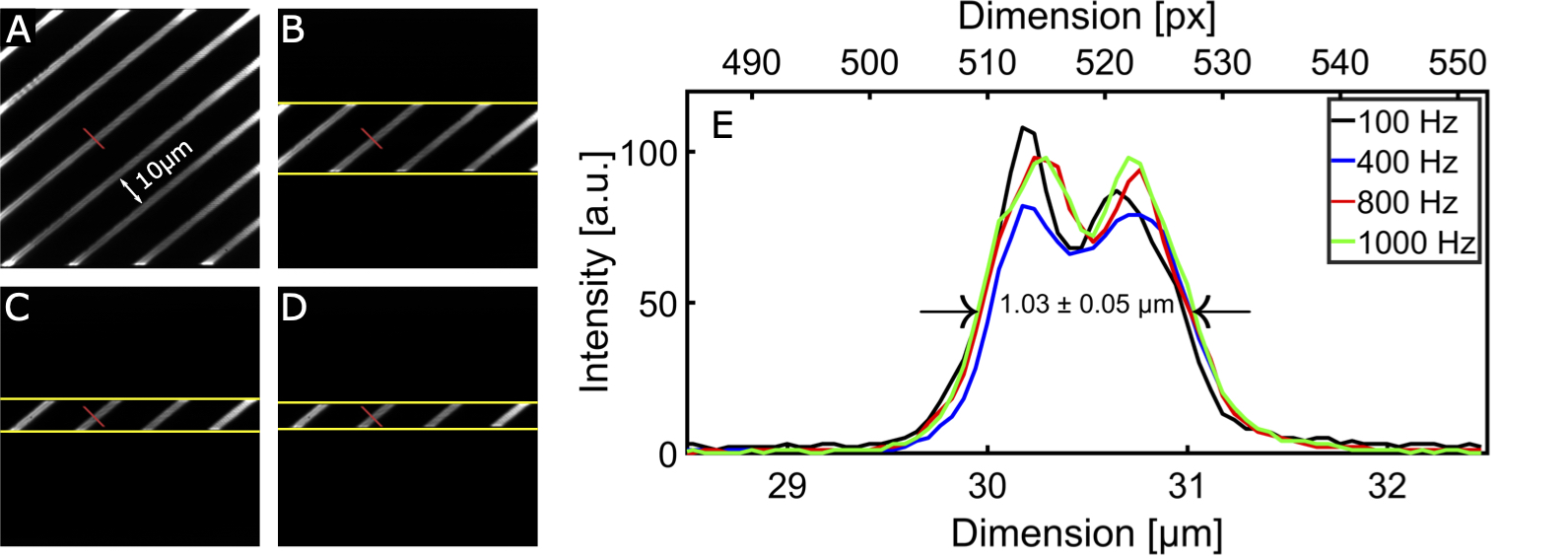}
	\caption{The resolution of the microscope is constant as a function of frame rate. (A) to (D) give the same area of a reflective line grid with a $10 \mu m$ spacing between the lines and $1 \mu m$ thick lines at different frame rates. The thickness of the grid line is correctly recorded for all frame rates and the averaged vale at the full width of half maximum is $1.03 \pm 0.05 \mu m $. The yellow lines mark the reduced width of the field of view at high frame rates. The superposition of the line profiles along the red line in (A) to (D) illustrates the independence of the lateral resolution on the frame rate (E). The lateral resolution in the sample plane is always in the order of $0.45 \pm 0.09 \mu m$. We calculated this resolution by fitting a tangent hyperbolic function to the slopes of (E). We defined the resolution as the lateral distance between 10\% and 90\% amplitude value of the hyperbolic function.}
	\label{fig:res_vs_fps}
	\end{center}
\end{figure*}

\gka{Vielleicht es für den kritischen Referee hilfreich, wenn wir die Erklärung, warum wir zwei Methoden zur Bestimmung der Auflösung benutzen, schon zu Beginn des Abschnitts bringen. Dann können wir diesen Faden nach den Experimenten wieder aufnehmen und mit den theoretischen Vorhersagen vergleichen.}
A fast and simple test to determine the optical resolution of this setup is to image the reflective grid lines of a micrometer scale (Objektmikrometer 3, Präzsisionsoptik Gera GmbH, Germany).
The distance between the grid lines is $10 \unit{\mu m}$ and a grid line is $1 \unit{\mu m}$ wide (Fig.~\ref{fig:res_vs_fps}).
To obtain a homogeneous illumination of the sensor, the amplitude of the $y$-scanning has to be adjusted to the frame rate. 
At low frame rates ($\leq 100 \unit{Hz}$), the entire sensor can be used for imaging, giving a field of view of approximately $60 \unit{ \mu m} \times 60 \unit{\mu m}$.
With increasing frame rate, the spacing between the lines written on the sensor can not be increased beyond a certain threshold while keeping a homogenous illumination of the sensor.
At frame rates above 100 Hz, the amplitude of the $y$-scanning has to be reduced, i.e., 320 lines were necessary for a homogeneous illumination of the $1024 \unit{px}\times 1024 \unit{px}$ sensor.
From this, we conclude that the spot size of the rescanned image is about 3 pixels in width (see also discussion of Fig.~\ref{fig:point-size} below).
In the series of images given in Fig.~\ref{fig:res_vs_fps}, for all frame rates $\ge 100 \unit{Hz}$  we keep all imaging parameters (laser intensity, scanning amplitude of the resonant mirrors, i.e., $x$-axis, etc.) constant.
However, we adjust the amplitude of the galvanic scanners ($y$-axis) to hold the density of lines on the sensor constant.
The image quality does not vary as a function of the frame rate, as can be seen from the superposition of the line profiles at the lowest and highest frame rate.
But the width of the field of view is reduced inversely to the frame rate, leaving only a small band of about $5.4 \unit{\mu}$m wide at 1000 Hz.
The reason for the douple peaks of the intensity profiles is that the scanned lines do not completely overlap each other.

Even though the field of view reduces with increasing frequency, the thickness of the grid line is still correctly depicted for all frame rates.
This is proven by extracting the averaged value for the full width of half maximum from the recorded intensity profile to be $1.03 \pm 0.05 \unit{\mu m} $ (Fig.~\ref{fig:res_vs_fps} E).
In addition, the lateral resolution of the microscope is gained by analyzing the edge steepness of the intensity profile. 
Therefore the slope of the measured intensity is fitted with a hyperbolic tangent.
The extent at which the amplitude value of the hyperbolic tangent declines from 90\% to 10\% is defined as the lateral resolution.
According to this our devices has a lateral resolution in the order of $0.45 \pm 0.09 \unit{\mu m}$. 

\begin{figure}[tbp]
	\begin{center}
	\includegraphics[width=0.8\columnwidth]{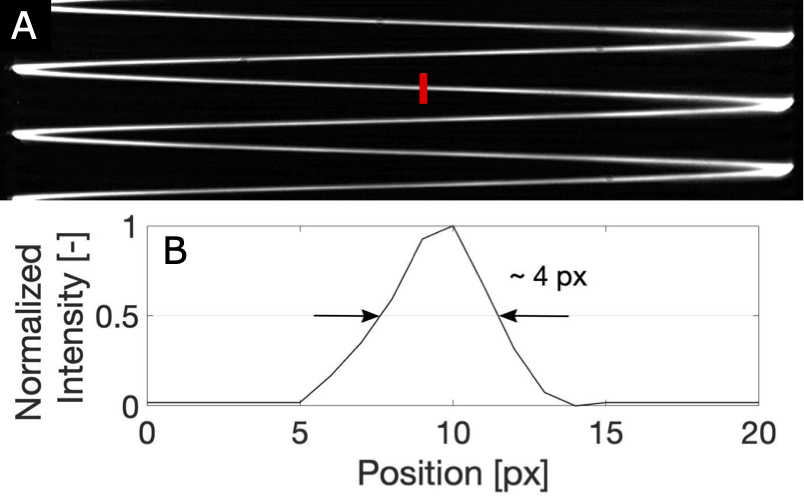}
	\caption{To visualise the width of the lines written on the 2D sensor we use a frame rate of 800 Hz and an amplitude for the $y$-scanning that leaves space between the scanned lines. The reflection of a completely optically reflective surface gives well separated lines (A) with a full width at half maximum of approximately $4 \pm 0.5$ pixel of the sensor (B).}
	\label{fig:point-size}
	\end{center}
\end{figure}

An alternative way to determine the resolution is the use of the line width of the rescanned beam on the sensor as an indicator for lateral resolution.
To separate the lines we use a high amplitude for the scanning along the $y$-axis at a high frame rate.
Hence, the lines are clearly imaged (Fig.~\ref{fig:point-size}).
We use a completely reflective optical surface of a universal calibration grid (PS20, Pyser Optics, Kent, UK).
Consistently with the above estimation, the full width at half maximum of the scanned line is in the order of $4 \pm 0.5$ pixel on the sensor (Fig.~\ref{fig:point-size}).
The magnification, meaning the transition from pixel to micrometer, is determined with calibration grid lines of the universal calibration grid to $17 \pm 1.1$ pixel$/ \unit{\mu m}$.
The magnification is constant for all frame rates.
While the magnification is constant for all frame rates, the experimental lateral resolution is calculated with the laser beam resolution ($4 \pm 0.5$ pixel) and the magnification ($17 \pm 1.1$ pixel$ / \unit{\mu m}$) to $0.235 \pm 0.034 \unit{\mu m}$.\\

Both methods of determining the resolution lead to different results. 
{\blue
The first method strongly depends on the quality of the grid lines and the chosen limits of the intensity drop.
This means that the resolution strongly depends on the quality of the measurement probe and should be recalculated for each measurement system.
The calculation of the resolution using the line width of the laser beam is independent of the sample and delivers the best possible experimental resolution.
}
%{\green
%As the first method strongly depends on the quality of the grid lines and the chosen limits of the intensity drop, determining the resolution with the line width is more coinciding with actual experimental resolution.
%}
The formula for lateral resolution \cite{egelman2012book}
\begin{equation}
r_{lat} = 0.51  \frac{\lambda}{NA}
\end{equation}
estimates the optimum lateral resolution to be $0.201 \unit{\mu m}$, when using a wavelength ($\lambda$) of 532nm and  numerical aperture ($NA$) of our objective $NA = 1.35$.
Regarding experimentally extracted value of $0.235 \unit{\mu m}$ the system nearly reaches theoretical predicted limit. 
The small deviation is explainable due to imperfections of the optical system. 
For example, the resonant mirrors are smaller than the back aperture of the objective. 
For this reason we are not able to illuminate the back aperture of the objective to its full extent. 
This decreases the actually $NA$ compared to the theoretical calculations causing experimental resolution loss.  

To complete this discussion about the resolution, the formula for the axial resolution \cite{egelman2012book}
\begin{equation}
r_{ax} = 0.88 \frac{\lambda}{n-\sqrt{n^2 - NA^2}}
\label{eq:ax_resolution}
\end{equation}
estimates the axial resolution of $0.568 \unit{\mu m}$, when using an immersion oil with an index of refraction ($n$) of 1.518. 
This theoretical axial resolution is worse than the theoretical lateral resolution by a factor of 2.8.
Consequently, a lower axial resolution is expected for confocal microscopes.
An important point is that the axial resolution depends on the squared NA of the objective.
Therefore the axial resolution is even more affected than the lateral resolution by a decrease of the NA which is true for the present setup (see section C below).

%Both shown experimental ways to calculate the lateral resolution are valid but show different results.
%Since the calculation with the edge steepness is more conservative and can be used for objects with unknown sizes we use this method in later sections.

\subsection{Bleaching of the sample}
Assuming a triangular motion of the resonant scanner, the dwell time can be estimated. 
At a line frequency of 32 kHz and a field of view of $1024 \unit{px} \times 1024 \unit{px}$ it amounted to about 122 ns to scan an area of the size of the optical resolution ($\approx 4 \unit{px} \approx 0.043 \unit{\mu m}^2$).
% \begin{align}
% d=\frac{1}{32000 \frac{\unit{lines}}{\unit{s}}}*\frac{\frac{4\unit{px}}{\unit{resolution \ sized \ spot}}}{1024 \frac{\unit{px}}{\unit{line}}}= \frac{122\unit{ns}}{\unit{resolution \ sized \ spot}}
% \end{align}
%This dwell time is significantly shorter than in spinning disk setups, where the dwell time is in the order of the inverse frame rate, i.e., in the millisecond range.
This dwell time is significantly shorter than in spinning disk setups, where the dwell time is in the microsecond range.
%\gka{Naja nicht so ganz, denn es gibt ja auch noch die Spotgrö\ss e. In einer Spinning disc fährt der Spot einmal pro Bild über eine Zeile, bei und über das ganze Bild. Die Dwell Time ist bei uns immer noch kürzer als bei Spinning Disc, aber der Unterschied ist nicht so ausgeprägt. Ich würde eine Dwell-Time von ungefähr 1/(Bildrate * Anzahl der Spots pro Zeile) erwarten. Bei 100 Hz und 1000 Spots pro Zeile sind das 10 $\mu$s. }
Even most single beam confocal microscopes working with single photon counting have higher dwell times, typically $ \ge 1 \unit{\mu s}$.
The short dwell times are known to reduce bleaching and photo toxicity\cite{graef2005}.
Despite the high intensity in the focus of the microscope objective ($ \approx 11.04 mW/ \unit{\mu m^2}$), we can image standard fluorescent samples (silica particles labelled with rhodamine, see next section) \cite{wenzl2013, Roth11_arx} more than 8000 times (xy-scan) without a significant amount of beaching. 
In earlier experiments \cite{wenzl2013,Roth11_arx} with the described confocal microscope and the used intensities ($ \approx 0.09 \unit{\mu W/ \mu m^2}$) we were only able to scan the same spot around 600 times due to a dwell time  of at least $ \approx 5 \unit{\mu s}$.

\subsection{3D sample of sedimented colloids}
\label{sect:sed_colloid}
\begin{figure}[tbp]
	\begin{center}
  \includegraphics[width=1\columnwidth]{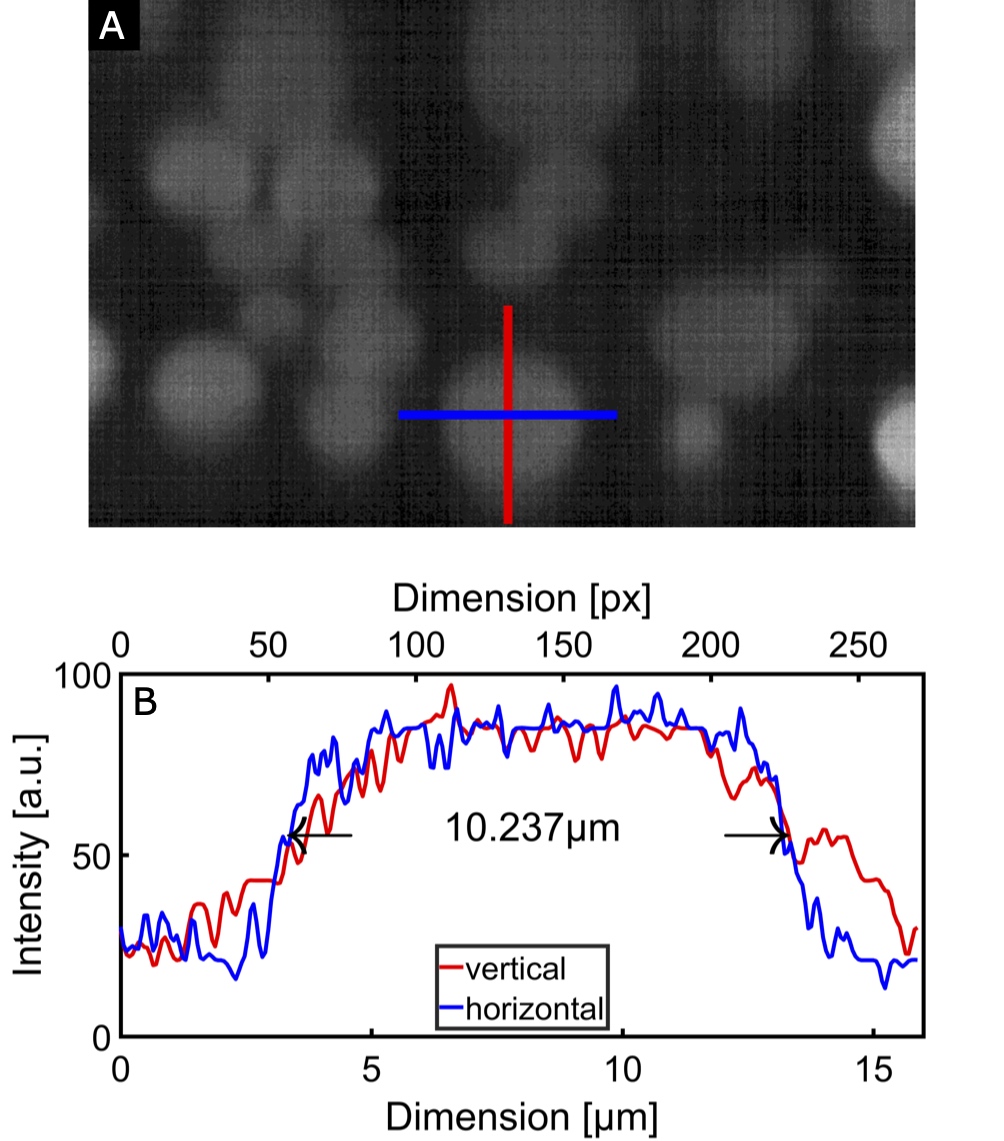}
	\caption{
	{\blue
	Image and analysis of a 3D sample of sedimented fluorescent colloids with a mean diameter of  $10 \unit{\mu m}$.
	The colloids are imaged with a 0.4Hz 3D stack rate and a 800Hz 2D frame rate.
	(A) shows a xz-cut through the measured 3D volume and (B) shows the intensity values along the blue (lateral or x-direction) and red (z-direction, vertical to the sample plane) line.
	The lateral and vertical size of the marked colloid ($\approx 10 \unit{\mu m}$) is identical in both directions. 
	}
	 %The resolution differs in the spatial directions.
	%The resolution in lateral direction (blue line) is in the order of $1.25 \unit{\mu m}$ (calculated from $\tanh$-fits to the intensity profile, compare supplementary material).
	%The vertical resolution is in the order of $4.1 \unit{\mu m}$.
	}
	\label{fig:xyz-resolution}
	 	\end{center}
\end{figure}

{\blue
\gka{Ich glaube der Titel dieses Unterabschnitts ist verwirrend. Es geht mehr um die Abbildung, denn um die Proben. Vielleicht: ''Characterizing the 3D imaging mode'' Dann kann im ersten Abschnitt immer noch gesagt werden, dass wir mit Blick auf die geplante Anwendung, Kolloide als Standardprobe benutzen. }
The new system was developed to measure the internal dynamic of dense colloidal systems.
\gka{Der Satz muss im letzten Absatz der Introduction stehen, vgl. mein Kommentar dort. Hier kommt dann eher was von Stil: ''Jetzt wollen wir zeigen, dass unser System tatsächlich den Anforderungen genügt.''}
Figure \ref{fig:xyz-resolution} A shows an xz-cut through a measured 3D volume of a dense colloidal probe.
The colloids are labelled silica particles with a mean diameter of $10 \unit{\mu m}$ and standard deviation of the size distribution of $1.7 \unit{\mu m}$.
Further details about the sample preparation are given in \cite{wenzl2013,Roth11_arx}.
The colloids are dispersed in an index matching solution of NaSCN in water.
The mismatch of refractive-index has to be lower than 0.005.
\gka{In absoluten Zahlen des Brechungsindex, oder relative Unterschiede. Ist zwar fast egal, sollte aber trotzdem präzise sein. }
{\green Otherwise scattering and defraction within the first layers of the sample degrade the resolution to a level that it is no longer possible to scan different layers of the colloid sample.}
Further information about the influence of a mismatch in refractive-index can be found in \cite{diaspro_02}.
Due to gravity, the particles sediment to the bottom of the measurement cell.
\gka{Was macht dieser Satz mitten zwischen den optischen Argumenten? Weiter nach oben, gleich nach '' . . . solution of NaSCN in water.''}
We image the colloids with a constant 3D stack rate of 0.4 Hz and a 800 Hz 2D frame rate (Fig.~\ref{fig:xyz-resolution}).
The $z$-motion of the focus is realized by moving the objective with {\green the} piezo stage.

As already mentioned, multi-beam scanning confocal microscopes lead to high background noise in dense samples with a high concentration of fluorescent objects.
Therefore, only single beam confocal microscopes are suitable to measure dense probes.
In earlier studies \cite{wenzl2013,Roth11_arx} the shearing of dense colloidal probes was studied. 
However, due to the small available scanning speeds only slow processes could be imaged.
To enable the measurement of fast dynamic processes in dense probes we developed the here described confocal microscope.
\gka{Hier brauchst du aber lange, um zu erklären, warum der Abschnitt wichtig ist. Ich würde nur erwähne, dass wir die Grenzen der Zeitauflösung für xz-Aufnahmen getestet haben. Da der kritische Referee hier eingehakt hat sollten wir wenig Schaum schlagen und vor allem Fakten erwähnen. Haben wir Bilder, die wir in die Supplementary Information packen könnten (auch wenn sie nicht besonders schön sind)? }
The new setup is able to scan an arbitrary xy-plane with a frame rate up to 1 kHz. 
A xz-plane can be recorded up to 200 Hz with a z-amplitude of $100 \unit{\mu m}$. 
Faster scan rates can be reached by lowering the z-amplitude. 
A xyz-volume (with a z-amplitude of $100 \unit{\mu m}$) can be recorded with a maximum 3D stack rate of around 200 Hz (with a xy frame rate of 160 Hz) .

{\green We already showed that the resolution depends strongly on the quality of the probe. 
This also holds for the colloidal sample \autoref{fig:xyz-resolution} A.}
The lateral and vertical size of the marked colloid ($\approx 10 \unit{\mu m}$) is identical in {\green the x and y direction.} 
Again we use the edge steepness as a criterion for resolution and fit a hyperbolic function (compare Supplementary Material for details) to the slopes of Fig.~\ref{fig:xyz-resolution} B.
The lateral resolution is in the order of $1.25 \unit{\mu m}$ and the vertical resolution is around $4.1 \unit{\mu m}$. 
{\green The measured axial resolution is increased by a factor of 3.28 relative to the lateral resolution.}
This is more than the theoretically predicted factor of 2.8.
The reason for that is that the axial resolution depends on the squared NA as already discussed in the previous section.
Therefore the axial resolution is even more decreased by a not fully used NA than the lateral resolution.
\gka{Die Argumentation verstehe ich nicht ganz. Vielleicht einfach den Vergleich mit den theoretischen Faktor (''This is more than the theoretically . . . '') weglassen.  }

{\green The measured lateral resolution is increased compared to the resolution we determined with Fig.~\ref{fig:res_vs_fps}.}
This is expected for the following reasons.
First, we no longer use a perfect reflective calibration grid.
The distribution of active dye inside the spheres cannot be determined independently and might show gradients close to the surface. 
The particles have a finite roughness on the surface. 
Despite the index-matching solution around the particles, minor imperfections in the index-matching would lead to refraction effects in the surface of the spherical particles. 
These refraction effects basically reduce the measured optical resolution. 
Furthermore, the fluorescent signal is weaker than the reflected signal leading to a reduced count rate.
This causes a low signal to noise ratio and therefore a reduced resolution.
\gka{Die Argumentation sollte die Motivation, warum auf verschiedene Arten die Auflösung bestimmen muss, wieder aufnehmen. }
}
\subsection{Possible improvements to the setup}
The analysis of the optical properties of the setup is based on several design aspects that were held constant for the sake of simplicity, but can be optimized.
To obtain optimal resolution, a good compromise between the beam expansion and the scanning angle of the resonant mirrors should be found.
A full illumination of the aperture by using bigger resonant mirrors would would enhance the resolution.
Larger mirrors however limit the resonance frequency of the mirrors and thus the possible maximum frame rate. 
Alternatively, increasing the beam  size between the sample scan unit and the imaging optics through including additional optical elements would result in the same. 
However, this requires larger scan angles in the sample scan unit, which are limited by design of the mirrors. 
To minimize negative effects of stray light the setup is completely enclosed. 
The camera works with an uncooled sensor
A cooled sensor would allow working with a lower dark count rate and thus with a higher signal-to-noise ratio.
In our realization, we use a camera with a relatively small sensor for practical reasons (pixel size: $13.9$ x $13.9 \unit{\mu m ^2 }$).
This leads to a diameter of the spot of about 4px. 
{\green For an optimal usage of the 2D sensor, the pixel size in the sensor should be adapted to the spot size, i.e., the spot size should be one pixel or a little larger.}
Smaller pixels lead to oversampling, i.e., not every pixel carries independent information. 
{\green An optimal choice of pixel size in the sensor enhances the signal} to noise ratio and thus the sensitivity of the entire setup. 

{\blue
{\red Hier weiter machen}
The ability to scan fast three-dimensional volumes of the setup also needs further improvements.
So far we did not include a special method to enable fast axial scanning like an electric or liquid lens.
At the moment we are using a piezo stage to move the objective up and down. 
This stage enables scanning speeds up to several hundred of Hz. 
However, due to the mechanical coupling of the immersion oil between microscope objective and sample, a fast movement of the objective distorts the sample.
\gka{Hierfür sollte ich Daten mit den Nanoindenter haben. Muss noch suchen. }
Therefore, we only scanned with slow axial speeds so far.
If we use an air objective this coupling would not exist but the resolution would further decrease.\\
To circumvent this problem other setups use e.g. liquid lenses. \cite{duocastella_14, piazza_18}
The lenses are placed at the back aperture of the objective and alter the laser beam to scan different axial positions.
Thus, the microscope objective is not moved and no distortions of the sample occur.
These lenses enable scanning with several kilohertz.
The next step in the development of the present setup might be the use of liquid lenses.\\
In the following section we present a different approach to measure three-dimensional dynamics. 
Instead of scanning the whole volume, we add a second beam path to utilize a defocusing technique.
}

{\blue
\section{Extensions of the setup}
The presented setup can easily be changed or expanded due to the easy access to the beam path.
This easy access is only possible with homebuilt setups since commercial microscopes do not offer these possibilities.
For example it is possible to do polarization measurements.
A polarization  prims can be included into the beam path and then it would be possible to do measurements with line scan rates of 32kHz.
It is also possible to include a remote focus system like describes in \cite{curran14} or liquid lenses. \cite{duocastella_14, piazza_18}
Furthermore it is possible to expand the setup with an additional beam path.
This second beam path corresponds to a standard fluorescent microscope by using the light coming from the objective before it enters the scanning units and the pinhole.
This opens the possibility to combine high-speed confocal imaging (standard beam path) with other fluorescent techniques (second beam path).

\subsection{Combination of confocal microscopy and astigmatism particle tracking velocimetry}
Here, we demonstrate the combination of the confocal imaging with a fluorescent method: astigmatism particle tracking velocimetry (APTV).
APTV is a three-dimensional velocimetry technique without the need to move the objective. \cite{kao94, cierpka11, rossi14}
It uses planned aberrations of the optical system to calculate the in plane position of fluorescent particles.
Therefore, the measurement speed is only a function of the xy-frame rate since two-dimensional information is used to generate the third (in plane) dimension.
Thus, the described setup can easily be used as an APTV device to measure three-dimensional dynamics, while keeping the simultaneous high-speed confocal imaging. 
The key idea is to use different optical wavelength for the different detections channels. 
In the following, we use the fluorescent light (shifted in wavelength) in the APTV channel and the reflected light (at the wavelength of the laser) in the high-speed confocal channel. 
With the same setup, using suitable filters and dichroic mirrors, one could also use fluorescent light of different wavelength in the two channels.
}
%{\green
%Normal Fluorescent imaging and astigmatism particle tracking velocimetry (APTV)\\
%The described setup can easily be changed to a standard fluorescent microscope by using the light coming from the objective before it enters the scanning units and the pinhole.
%This opens the possibility to combine high-speed confocal imaging with other fluorescent techniques. 
%Here, we demonstrate normal fluorescent imaging on an additional camera and its extension by an astigmatic aberration.
%Thus, the described setup can easily be used as an astigmatism particle tracking velocimetry (APTV) device to measure three dimensional dynamics, while keeping the simultaneous high-speed confocal imaging. 
%The key idea is to use different optical wavelength for the different detections channels. 
%In the following, we use the fluorescent light (shifted in wavelength) in the APTV channel and the reflected light (at the wavelength of the laser) in the high-speed confocal channel. 
%With the same setup, using suitable filters and dichroic mirrors, one could also use fluorescent light of different wavelength in the two channels. 
%}
\begin{figure}[tbp]
 \begin{center}
  \includegraphics[width=1\columnwidth]{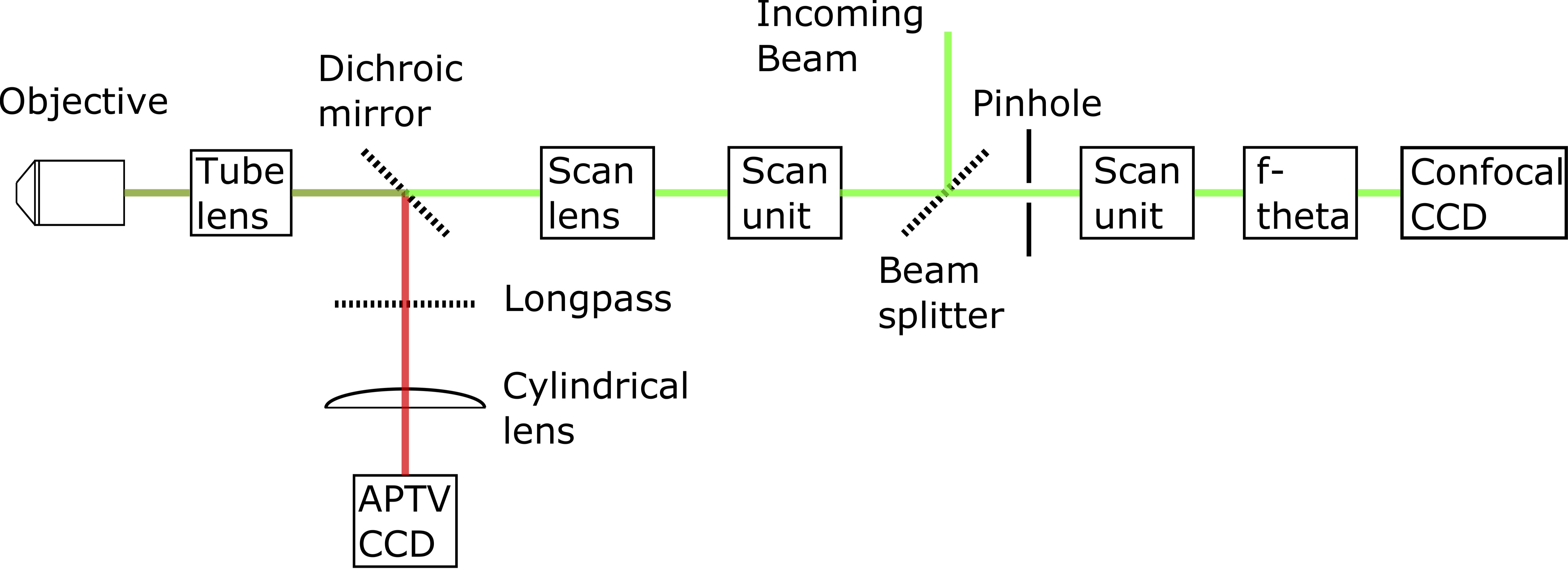}
 \caption{Schematic of the astigmatic beam path in parallel to the high-speed confocal imaging. 
 The fluorescent light for the APTV channel is reflected by a second dichroic mirror towards the cylindrical lens which creates an astigmatic aberration. 
 This deformed image is recorded with the CCD camera. 
 The shape information of tracer particles can be used to calculate the in plane position of the particles. 
 The reflected light for the high-speed confocal channel is detected by the camera of our confocal microscope. 
 Therefore, this setup enables to capture a high-resolution high-speed confocal image synchronized with an astigmatic fluorescent image.}
 \label{fig:Schematic_Astigmatic}
\end{center}
\end{figure}

To do so, an additional dichroic mirror (DMSP550, Thorlabs, Munich, Germany) and a longpass filter (Qioptiq, 550nm, Germany) were placed between the objective and scan lens in the imaging optics (\autoref{fig:Schematic_Astigmatic}).
The dichroic mirror reflects the fluorescent light to the APTV CCD (a second camera, CS2100M-USB, Thorlabs, Munich, Germany) which is synchronized with the confocal CCD (standard CCD camera of the setup).
The additional longpass filter is necessary to eliminate all remaining laser light in this channel.
Since no scanning unit and no pinhole are located in the beam path in front of the APTV CCD, this part of the setup equals a fluorescent microscope.
The reflected light passes through the additional dichroic mirror and follows the standard light path of our confocal microscope.
Hence, the setup delivers a confocal reflexion image simultaneous to a fluorescent microscope image.

\begin{figure}[tbp]
 \begin{center}
\includegraphics[width=83mm]{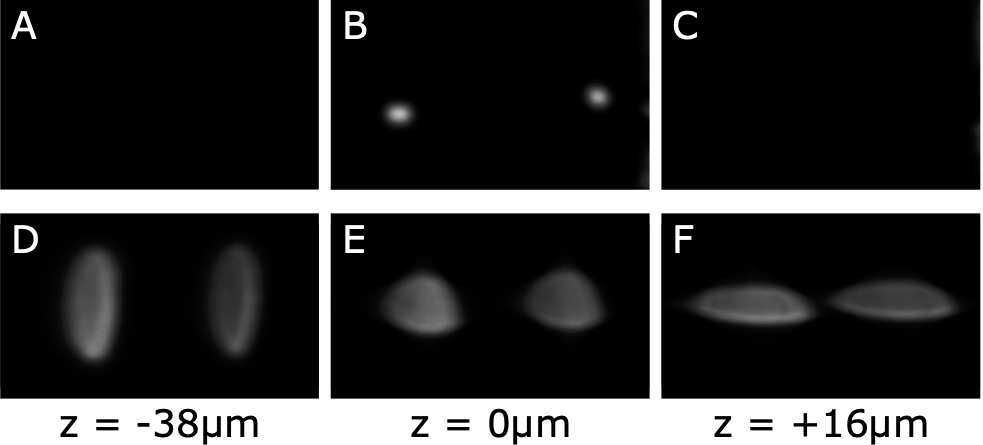}
 \caption{Comparison of confocal (top) and APTV (bottom) images of 4 $ \unit{\mu m}$ diameter fluorescent particles for different focus heights (z). 
 In the focus position ($z = 0 \unit{\mu m}$) the confocal  image shows nice circular shapes of the particles. 
 The APTV image shows already slightly deformed circular shapes. 
 In the case of out of focus postions ($z = -38 \unit{\mu m}$, $z = +16 \unit{\mu m}$) the confocal microscope cannot record the particles anymore since all out of focus informations are blocked by the pinhole. 
 Whereas the APTV images clearly show the particles. 
 The images of the particles are deformed to clearly distinguishable ellipsoidal shapes. 
 The shape information can be used to calculate the in plane position of the particles.}
\label{fig:Conf_Astigmatic_Comp}
\end{center}
\end{figure}
 
By placing a cylindrical lens (f = 50mm, Thorlabs, Munich, Germany) in front of the APTV CCD,  astigmatic aberrations are created in the images.
The fluorescent light of the spherical particles is distorted such that they are depicted elliptical in the sensor image. 
This is best illustrated by comparing confocal images with APTV images at different focus planes (\ref{fig:Conf_Astigmatic_Comp}). 
Therefore,  two $4 \unit{\mu m} $ fluorescent particles (Microparticles, Germany) are imaged for different objective positions (20X objective, NA = 0.45, Olympus, Japan). 
Starting with the focus position in the particle plane ($z = 0 \unit{\mu m}$) confocal image (B) shows two spherical particles whereas the APTV image (E) displays slightly deformed circular shapes. 
If the focus position is changed to lay underneath the particle plane ($z = -38 \unit{\mu m}$) no fluorescent light is recorded with the confocal mode as the pinhole blocks all light coming from the particles (A). 
In comparison the APTV image shows two ellipsoidally deformed particles (D). 
Changing the focus to lay above the particle plane ($z = 16 \unit{\mu m}$) results in the same black confocal image (C) as before due to known reasons. 
But again to a distortion to ellipsoids in the astigmatic beam path (F).
Note the opposite elongation of the ellipses compared to the images below the focal plane. 
The deformation of fluorescent tracer particles in the APTV images can be used to extract the $z$-position of the particles in relation to a predefined focus plane. \cite{kao94, cierpka11, rossi14}\\

{\blue
\subsection{Dynamic wetting behavior of droplets}

 \begin{figure}[tbp]
 \begin{center}
  \includegraphics[width=82.5mm]{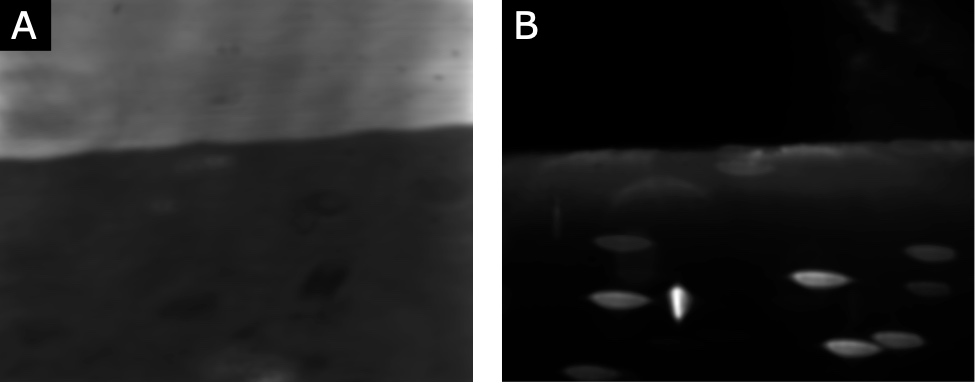} 
 \caption{Synchronized confocal (A) and astigmatic (B) measurements of a moving three phase contact line. 
 Polystyrene tracer particles are dispersed in a density matching solution of pure water and deuterium oxide.  
 In the confocal reflection, the contact line is clearly visible as the sharp contrast in the reflected intensity. 
 In the astigmatic picture the deformed tracer particles are clearly visible. 
 APTV tracking algorithms can be used to track the particles in three dimensions. 
 The in plane position is provided by a shape analysis of the particles.}
 \label{fig:Astigmatic}
    \end{center}
\end{figure}

We could use this configuration also to measure the dynamic of dense systems. 
Instead of labelling all colloids with the same dye, we could use different dyes for the different detection paths (APTV and confocal). 
APTV could track the three-dimensional movement of some specially labelled colloids while the confocal scans a specific plane through the sample.
This would provide a three-dimensional deformation field of the dense system.\\
Here, we want to highlight that this configuration is also useful to explore wetting phenomena.
Most studies on dynamic wetting behavior are macroscopic and measure the contact angle and the spreading behavior to quantify the results.
Only a few studies looked at the internal dynamics of liquids during wetting or dewetting processes due to the lack of suitable velocimetry techniques.
With APTV it is possible to measure the three-dimensional dynamics with a good resolution.
Here, we are able to combine an APTV measurement with a high resolution confocal measurement.\\
}
As an example of the possibilities of such a combined detection, we choose the motion of a contact line (\autoref{fig:Astigmatic}). 
In such  wetting experiments, some of the interesting quantities are the position and dynamics of the contact line and the flow field inside the wetting liquid. \cite{moffatt64, huh71, voinov76, cox86, henrich_16}
The setup described here allows measuring both quantities with the highest possible resolution simultaneously. 
The reflected light of the substrate allows a precise detection of the contact line, because the reflected intensity changes strongly at the contact line. 
The fluorescent light of tracer particles suspended in the wetting liquid, gives information on the 3D position (and trajectory) of the tracer particle, and thus the flow field. 
Fluorescent polystyrene particles with a size of $4 \unit{\mu m}$ are dispersed in the measurement solution. 
To avoid sedimentation of the particles a 1:1 mixture of pure water (Satorius: Arium\textregistered pro VF/UF \& DI UV with a resistivity of \mbox{18.2 $M \Omega $)} and deuterium oxide (Sigma-Aldrich 99\%) was used.
The confocal reflexion image (A) and the astigmatic fluorescent image (B) show the same imaged area at the same time.
In the confocal picture (A) we can clearly see the contact line but no tracer partciles since none are in the focal plane at this specific moment.
In the astigmatic picture (B), we see different elliptical shapes which are image particles distorted by the aberration of the cylindrical lens.
With particle tracking software the $x$- and $y$-coordinates can be detected. \cite{kao94, cierpka11, rossi14}
Using a shape analysis of the particles one can calculate the third (in plane) coordinate. \cite{kao94, cierpka11, rossi14}
In this configuration horizontally deformed shapes indicate that the particles are close to the cover slip.
The vertically deformed shape indicates that this particle is further above the cover slip. 
Therefore, this configuration allows to measure a confocal reflexion image synchronized with an astigmatic three-dimensional particle tracking device.\\

Further options and applications are now open with this design. 
The reflectance channel could be replaced with a second or multiple other fluorescent channels if a second or more laser light sources are used in the setup. 
Therefore, structures and tracer particles could be labelled with fluorescent dye leading to more flexible experiments showing high resolution confocal structures and three dimensional dynamics at the same time.
%{\blue
%Whereas in standard APTV it is only possible to scan a xy-plane and calculate the z coordinate we can also scan a xz-plane and calculate the y coordinate. 
%We can freely choose which plane is scanned and which coordinate is calculated. 
%Of course it is also possible to scan an arbitrary plane with changing xyz-coordinates and calculate the coordinate which is perpendicular to it.
%This is only possible because we can scan any possible plane.
%}

\section{Conclusions: Advantages, perspectives and limits}

In this work we have presented a novel setup for fast scanning confocal microscopy.
This setup uses only a single scanning laser beam and a rescanning unit to project the information from the sample directly on a 2D sensor.
The resolution of the setup is independent of the frame rate.
The lateral resolution is  $0.235 \unit{\mu m}$ in an ideal configuration (reflection from a calibration grid).
{\blue
For dense colloidal probe samples the lateral resolution is in the order of $1.25 \unit{\mu m}$ and the axial resolution is in the order of $4.1 \unit{\mu m}$.
}
The reason for this difference in resolution is the reduced signal to noise ratio for the measured fluorescent particles.
By construction, cross-talk between scanning beams (like in spinning disk setups) is not possible in our setup.
The imaging speed is limited by the frequency of the fast (resonant) scanning mirrors (here 32 kHz line frequency).
Further increase in scanning speed necessitates faster scanning units that allow for a high beam diameter.
Scanning a single beam with high speed over the sample leads to short dwell times on each spot of the sample (here about 122 ns) and reduces bleaching and phototoxicity.
This short dwell time compensates the needed higher laser powers in comparison to other setups.
The resolution of the setup can be enhanced by a better camera with higher quantum efficiency and a pixel size that is equivalent to the spot size.

Since all three axis for scanning the sample in 3D can be driven individually, the plane projected on the 2D sensor can be chosen freely. 
Furthermore a vertical cut through the sample (an $xz$-plane) could be projected on the 2D sensor.
{\blue 
The highest frame rates are 1 kHz for xy-scans and 200Hz for xz-scans with a z-amplitude of $100 \unit{\mu m}$. 
Reducing the z-amplitude provides faster scanning.
An xyz-stack with a z-amplitude of $100 \unit{\mu m}$ can be recorded with 200 Hz (160 Hz xy frame rate).
}
This flexibility is only possible when using a single scanning beam and would be impossible in a mutli-beam setup.

The design of our setup allows introducing additional fluorescent imaging techniques along confocal imaging. 
We demonstrated this by separeting the normal fluorescent image from the confocal beam path using a dichroic mirror. 
This is then used to show the possibility for astigmatism particle tracking velocimetry. 
One way to measure 3D velocity fields is the described APTV which can be easily used in the setup. 
Therefore, this configuration allows to measure a confocal image synchronized with an astigmatic three-dimensional particle tracking device.

\section*{Supplementary material}
See supplementary material for details about the calculation of the resolution via edge steepness.

\section*{Acknowledgements}
The authors gratefully acknowledge financial support of the project through the DFG within the CRC 1194 (project A06).

\bibliography{Literatur}
\end{document}